\documentclass{aa}  

\usepackage{graphicx}
\usepackage{txfonts}
\usepackage[colorlinks=true, citecolor=blue, linkcolor=blue]{hyperref}
\usepackage{xcolor}
\usepackage[normalem]{ulem}

\usepackage{natbib}
\bibpunct{(}{)}{;}{a}{}{,} 

\newcommand{\mps}{m\,s$^{-1}$}
\newcommand{\Wmm}{W\,m$^{-2}$}

\begin{document}

   \title{IRIS observations of chromospheric heating by acoustic waves \\   in solar quiet and active regions}

   \subtitle{}

   \author{V. Abbasvand
          \inst{1,2}\fnmsep\thanks{vahid.abbasvand.azar@asu.cas.cz}
          \and
          M. Sobotka\inst{1}
           \and
          M. \v{S}vanda\inst{1,2}
           \and
          P. Heinzel\inst{1}
         \and
          W. Liu\inst{1} 
          \and
          L. Mravcov\'{a}\inst{1,2}
          }

   \institute{Astronomical Institute of the Czech Academy of Sciences (v.v.i.),
Fri\v{c}ova 298, 25165 Ond\v{r}ejov, Czech Republic
\and
 Astronomical Institute of Charles University, Faculty of Mathematics and Physics,
V Hole\v{s}ovi\v{c}k\'{a}ch 2, 180 00 Praha 8, \\ Czech Republic
 }

  \date{Received 14 January 2021; accepted 11 February 2021}


\abstract
   {}
   {To study the heating of solar chromospheric magnetic and nonmagnetic regions by acoustic and magnetoacoustic waves, the deposited acoustic-energy flux derived from observations of strong chromospheric lines is compared with the total integrated radiative losses. }
   {A set of 23 quiet-Sun and weak-plage regions were observed in the Mg\,\textsc{ii} k and h lines with the Interface Region Imaging Spectrograph (IRIS). The deposited acoustic-energy flux was derived from Doppler velocities observed at two different geometrical heights corresponding to the middle and upper chromosphere. A set of scaled nonlocal thermodynamic equilibrium 1D hydrostatic semi-empirical models ---obtained by fitting synthetic to observed line profiles--- was applied to compute the radiative losses. The characteristics of observed waves were studied by means of a wavelet analysis.}
   {Observed waves propagate upward at supersonic speed. In the quiet chromosphere, the deposited acoustic flux is sufficient to balance the radiative losses and maintain the semi-empirical temperatures in the layers under study. In the active-region chromosphere, the comparison shows that the contribution of acoustic-energy flux to the radiative losses is only 10--30\%.}
   {Acoustic and magnetoacoustic waves play an important role in the chromospheric heating, depositing a main part of their energy in the chromosphere. Acoustic waves compensate for a substantial fraction of the chromospheric radiative losses in quiet regions. In active regions, their contribution is too small to balance the radiative losses and the chromosphere has to be heated by other mechanisms.}

  \keywords{Sun: chromosphere -- Sun: oscillations -- Radiative transfer}
  \titlerunning{IRIS observations of chromospheric heating by acoustic waves}

\maketitle
%

\section{Introduction}

The rise of temperature in the solar chromosphere and corona implies additional heating mechanisms  in conjunction with radiative energy transport. There is no generally accepted picture of the energy transport processes responsible for this rise and pure radiative transfer cannot explain the temperature given by semi-empirical models.
Spectral lines typical for the chromosphere
are strong hydrogen lines, Ca\,\textsc{ii} K and H and infrared triplet, and the k and h lines of Mg \,\textsc{ii}. Cores of these spectral lines are formed in conditions that are far from local thermodynamical equilibrium (LTE). The radiative energy released in these lines from the quiet-Sun chromosphere is estimated to 4300~\Wmm{} ---excluding Lyman-$\alpha$ \citep{Vernazza1981}--- and is  greater in active regions.

The scenarios of chromospheric heating processes can be divided into two main competitive categories \citep[see ][for a review]{Jess2015}: (i) Reconnection of magnetic field lines with opposite directions leads to a conversion of magnetic to thermal energy \citep{Testa2014}; and (ii) propagation of waves in plasma: acoustic waves are generated by material motions in the upper convection zone and the energy of these waves is dissipated at some distance from the birth point because of the change in atmospheric properties along the vertical direction in a stratified medium \citep{Kayshap2018}.  

The magnetic field acts as a catalyst in the propagation of acoustic waves without being destroyed in the process. In particular, inclined magnetic field reduces the magnetic cut-off frequency 5.2 mHz \citep{Bel1977, Cally2006}, allowing the propagation of low-frequency waves into the chromosphere. Two different cases can be considered in quiet-Sun regions concerning the magnetic field topology: the more vertical field in solar magnetic network and the more inclined one in the intranetwork. Mostly vertical magnetic field lines in the photosphere can be more inclined in the chromosphere and expand in the atmospheric volume at some height. This field expansion leads to canopy regions, which substantially reduce the energy flux carried by the waves \citep[so-called magnetic shadows, ][]{Judge2001, Vecchio2007, Konto2010}.  

Magnesium is one of the most abundant elements in the solar atmosphere. The Mg\,\textsc{ii} k and h resonance lines are important tools to study the solar chromosphere, sampling layers from the temperature minimum to the upper chromosphere. The solar spectra of each line are often characterized by a central emission reversal surrounded by two emission peaks. 
The emission peaks form in the middle chromosphere, where a decoupling between their source function and the Planck function occurs. The central reversal forms at a height that is 200 km below the transition region, where the decoupling is even stronger. The k line is formed a few tens of kilometers higher in the atmosphere than the h line because its opacity is higher by a factor of two \citep{Leenaarts2013ApJ_formation}. These latter authors studied k and h line formation using a 3D radiation-magnetohydrodynamic simulation performed with the {\it Bifrost} code \citep{Gudiksen2011Bifrost}, which represents mainly quiet-Sun conditions rather than those in active regions. The formation heights of the line features are strongly model dependent.

A number of different solar atmospheric models derived from observations or theoretical simulations are used to study the physical properties of the solar atmosphere from the photosphere to the transition region. In the following, we refer to the set of semi-empirical  1D hydrostatic models, VAL A--F, presented by \cite{Vernazza1981}, which describe the solar atmosphere from intranetwork to bright network features. In their theoretical review, \cite{Ulmsch2003} claimed that acoustic waves can heat the intranetwork regions on the Sun. Many studies have been based on 3D heating simulations. \cite{Wedemeyer2007} studied whether or not acoustic waves are sufficient to heat the quiet-Sun intranetwork regions. They compared TRACE observational data with a synthetic image obtained from a 3D simulation and demonstrated that these observations of intranetwork regions can be reproduced without a strong need for magnetic fields. In a similar work, \cite{Cuntz2007} also showed that high-frequency acoustic waves can heat those regions in the solar chromosphere. Some recent works have shown that heating by waves partly balances the radiative losses of the chromospheric layers \citep{Jefferies2006,Kalkofen2007, Bello2010a, Kanoh2016, Grant2018, Abbasvand2020, Abbasvand2020AA}, while  other authors have declared that the total acoustic flux energy is still insufficient to heat the quiet-Sun regions \citep{Fossumetal2005, Carlsson2007, Beck2009}.

The dissipation of acoustic waves is a multi-dimensional time-dependent process and difficulties come from the modeling and its high computational requirements, which do not allow us to represent the observed chromospheric structures directly this way \citep{Carlsson2012}. A review of radiation-hydrodynamic models of the solar atmosphere was recently published by \cite{Leenaarts2020}.

In an alternative, stationary approach, time-averaged atmospheric parameters of 1D semi-empirical hydrostatic models were used to show that acoustic energy flux deposited in the chromosphere provides a remarkable source of energy to balance the local radiative losses \citep{sobotka2016}. Recently, using the same approach, \citet{Abbasvand2020} introduced a new grid of 1D semi-empirical hydrostatic models by scaling the temperature and column mass of six initial VAL A--F models by \cite{Vernazza1981} in order to refine the comparison of the deposited acoustic flux with radiative losses.

\cite{Abbasvand2020, Abbasvand2020AA} analyzed four observations of quiet-Sun and weak active regions in the lines Ca \textsc{ii} 8542 \AA , H$\alpha$, and H$\beta$. These latter authors demonstrated that the radiative losses can be fully balanced by the deposited acoustic energy flux in the middle chromosphere of a quiet Sun region sufficiently far from any extended canopy region. The acoustic energy flux was reduced by a factor of between two and three in the quiet-Sun region that was close to a plage and a pore. In the upper chromosphere, the contribution of the deposited acoustic flux energy to the released radiative energy was small, namely less than 20\% both in quiet and active regions. No clear conclusion was obtained for the middle chromosphere of active regions because of a lack of data.

The present work is based on observations of 12 quiet and 11 active regions acquired in the Mg\,\textsc{ii} h and k lines. We use this large data sample to verify the statistical significance of our previous results and to obtain conclusive information about the heating of active regions (plages). We also study the character of observed waves by means of a wavelet analysis.


\section{Observations and data processing} \label{sec:observation}

\begin{table*}\centering
\caption{List of IRIS observations.}  \label{tab:IRIS_table}
\begin{tabular}{llclcccc}
\hline\hline
  &  &   &  &  &  & \\
Data  & Date and Time & Slit  & OBSID & Raster Cadence & t$_{\rm exp}$ & No. of spectral scans\\
  &  &  position  &  & [s] & [s]&   \\

\hline

  &  &   &  &  &  &  \\
QS1  & 2017-05-22 07:31UT & 1  & 3633105426 & 24.7&  2 & 648 \\
QS2  & 2019-03-03 14:16UT & 1  & 3630109417 & 36.7&  8 & 400 \\
QS3  & 2019-05-17 08:54UT & 1  & 3620106017 & 20.8&  4 & 280 \\
QS4  & 2019-05-20 09:05UT & 1  & 3620106017 & 20.8&  4 & 200 \\
QS5  & 2019-05-25 08:44UT & 1  & 3620106017 & 20.8&  4 & 230 \\
QS6  & 2019-06-14 06:43UT & 1  & 3620106417 & 20.8&  4 & 380 \\
QS7  & 2019-06-15 11:13UT & 1  & 3620106417 & 20.8&  4 & 450 \\
QS8  & 2019-06-16 11:13UT & 1  & 3620106417 & 20.8&  4 & 450 \\
QS9  & 2019-06-19 08:09UT & 1  & 3620106417 & 20.8&  4 & 240 \\ 
QS10 & 2019-06-19 11:09UT & 1  & 3620106417 & 20.8&  4 & 420 \\
QS11 & 2019-07-06 07:30UT & 1  & 3660106017 & 20.8&  4 & 230 \\
QS12 & 2019-07-17 07:29UT & 1  & 3660106017 & 20.8&  4 & 175 \\
\hline
  &  &   &  &  &  & \\
AR1  & 2015-06-25 07:29UT & 1  & 3630105426 & 24.7& 4  & 199 \\
AR2  & 2015-07-08 22:16UT & 4  & 3660106120 & 20.8& 4  & 400 \\
AR3  & 2015-07-22 14:47UT & 1  & 3660108117 & 36.7& 8  & 431 \\
AR4  & 2015-07-22 14:47UT & 2  & 3660108117 & 36.7& 8  & 431 \\
AR5  & 2017-05-20 07:31UT & 1  & 3633105426 & 24.7& 2  & 630 \\
AR6  & 2017-05-20 07:31UT & 3  & 3633105426 & 24.7& 2  & 630 \\
AR7  & 2017-05-21 07:31UT & 1  & 3633105426 & 24.7& 2  & 300 \\
AR8  & 2017-05-21 07:31UT & 3  & 3633105426 & 24.7& 2  & 300 \\
AR9  & 2017-05-25 07:33UT & 2  & 3633105426 & 24.7& 2  & 430 \\
AR10 & 2019-06-12 07:01UT & 1  & 3620106417 & 20.8& 4  & 400 \\
AR11 & 2019-06-12 07:01UT & 4  & 3620106417 & 20.8& 4  & 400 \\
\hline
\end{tabular}
\tablefoot{Data labels are: QS for quiet Sun and AR for active region; t$_{\rm exp}$ stands for exposure time.}
\end{table*}

We studied 23 data sets obtained in the lines Mg\,\textsc{ii} k and h (2796.35 \AA~and 2803.52 \AA, respectively) by the Interface Region Imaging Spectrograph \citep[IRIS,][]{DePontieu2014SoPh}. Details of the data sets are presented in Table~\ref{tab:IRIS_table}. The targets were 12 quiet-Sun and 11 weak active regions located close to the center of the solar disk to minimize possible projection effects \citep{Ghosh2019}. IRIS provides high-resolution slit-jaw images as well as spectra of areas covered by the slit. These observations have a spatial sampling of 0.33$''$/pixel along the slit and 0.33$''$/pixel perpendicular to the slit with the exception of \mbox{OBSID} 3660106120, where the spatial sampling was 1$''$/pixel. The spectral pixel size was 51.2 m\AA. We used continuous series of spectral scans that were not contaminated by tracks of cosmic particles. Usually, only one slit position was used, but in four cases of active-region observations we included two positions to increase the statistical weight of our results. The IRIS slit-jaw images in the 2832~\AA ~pseudo-continuum passband were used to align the spectroscopic observations with magnetic-field maps. We note that in eight data sets (QS1, QS2, AR1, AR5, AR6, AR7, AR8, and AR9) the slit-jaw images were not recorded. 

We used IRIS Mg\,\textsc{ii} k and h calibrated Level 2 data, which have been processed for dark current subtraction, flat-field correction, geometrical correction, and wavelength calibration \citep{DePontieu2014SoPh, Wulser2018SoPh}. We then performed a radiometric calibration to convert the IRIS data, originally in units of data number (DN), into absolute intensity units (erg~s$^{-1}$~cm$^{-2}$~sr$^{-1}$~Hz$^{-1}$), applying the \texttt{iris\_calib\_spectrum} function, which is available in the IRIS package of the SolarSoft system \citep{FreelandHandy1998}. We used time-dependent effective areas (in cm$^2$) for the radiometric calibration by specifying the date and time of each observation.
For each position along the slit of each data set, the mean profile calibrated in absolute intensity units was obtained by time-averaging over the observing period. This was used to find the most appropriate local semi-empirical model.

\begin{figure}[t]\centering
\includegraphics[width=0.49\textwidth]{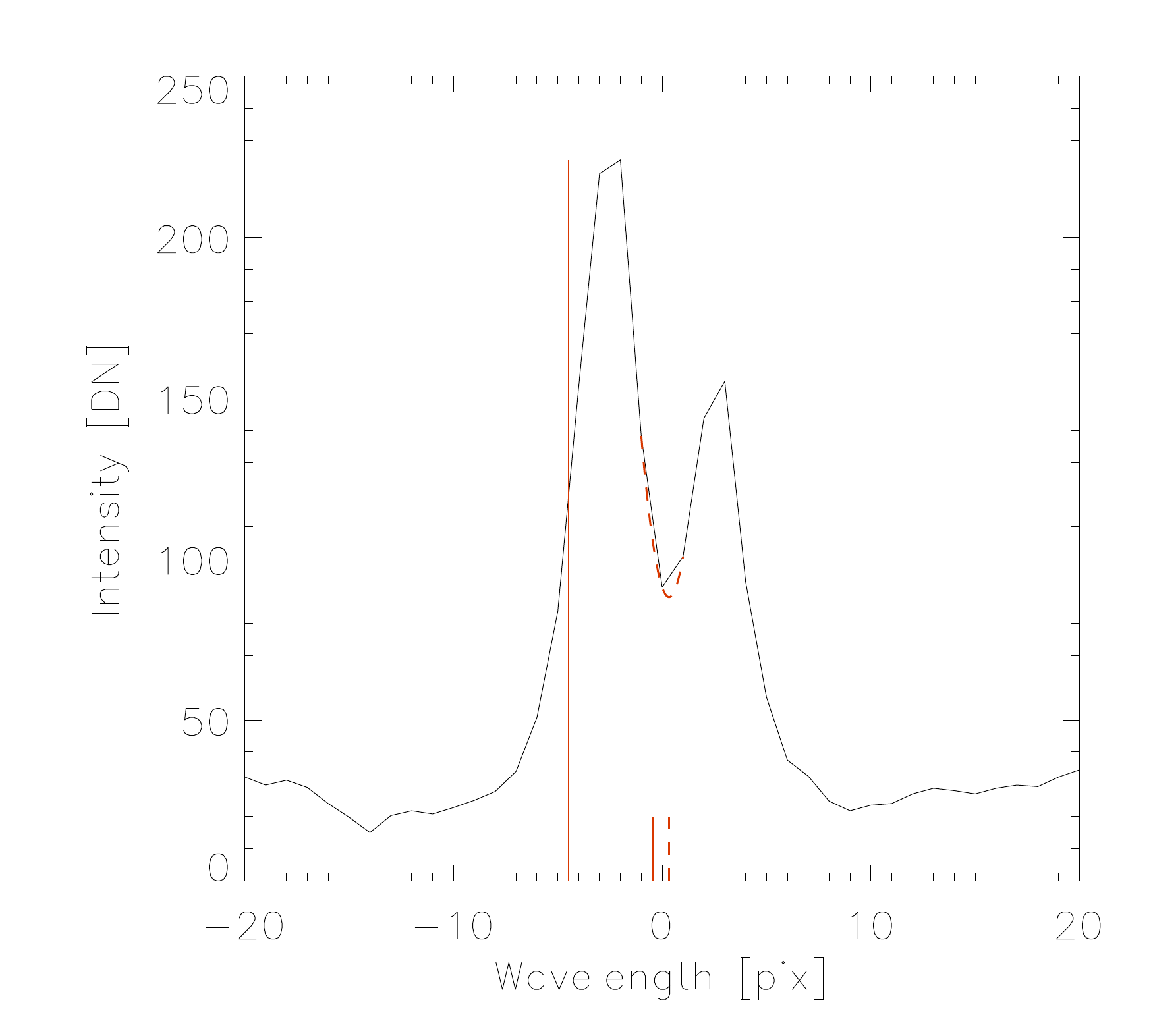}
\caption{Methods of Doppler-shift measurement in the observed profile (Mg\,\textsc{ii} k): parabolic fit for the central reversal and double-slit method for the emission core. Short red lines at the bottom show the shifts.\label{fig:dopmes}}
\end{figure}

Doppler shifts of the k and h emission and central reversal were measured to obtain chromospheric Doppler velocities required for the calculation of acoustic fluxes. An illustration of the measurement methods is shown in Fig.~\ref{fig:dopmes}. We looked for the positions of the central reversal and the emission core in a spectral region of 2.05~\AA ~(40 pixels) in width around the approximate line center. The central-reversal position was determined by means of a three-point parabolic fit of the profile minimum between the two emission peaks.
For those profiles where the central reversal was missing (single-peaked emission profiles), the central-reversal velocity was set to zero and marked as an outlier.

\begin{figure}
    \centering
    \includegraphics[width=0.49\textwidth]{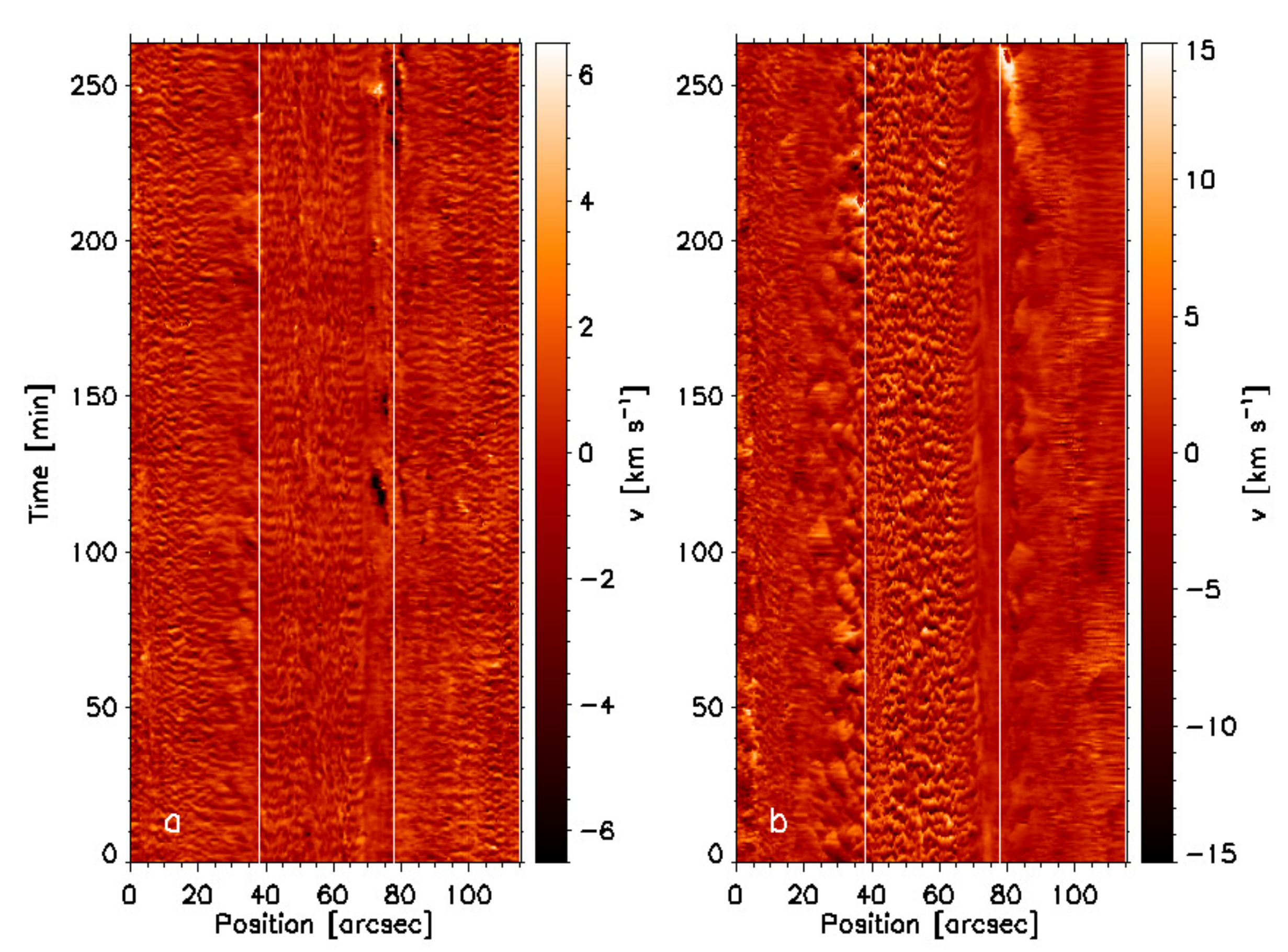}
\caption{Example of a Dopplergram for the data set AR4. Panel ({\it a)} shows  velocities of the emission core; Panel ({\it b}) shows velocities of the central reversal. White vertical lines delimit a magnetic region.}
    \label{fig:Dop20}
\end{figure}

Doppler shifts of the emission core, including both emission peaks, were measured by means of a ``double-slit'' method \citep{Garcia2010}. This method consists in the minimization of the difference in intensities of light passing through two ``slits'' located at a certain distance in the opposite wings of the emission core. The line profile is shifted by sub-pixel distances with respect to the slits to find the minimum intensity difference. The distance between the slits determines the depth of the profile, for which the shift is measured. We selected the slit distance to match the half intensity difference between the emission maximum and the line-profile minimum, so that shifts of the whole emission core were measured independently of the emission peaks and central reversal. Because the widths of the k and h emission cores change with physical conditions in the chromosphere, the slit distance was adapted to the emission-core width measured for each time instant and position.

The shifts were converted into Doppler velocities, defining the reference zero as the time-average of measurements for each position along the slit separately, and 2D time-position Dopplergrams were constructed. Outliers and failures of measurements were replaced by interpolated values obtained by median filtering ($5 \times 5$ pixels surroundings) of these Dopplergrams. Pairs of Dopplergrams for the k and h lines were obtained in this way. Because the Dopplergrams in the pair were always almost identical, their average (Fig.~\ref{fig:Dop20}) was used for the acoustic-flux calculation.

The photospheric magnetic-field strength and inclination are necessary for the acoustic-flux calculation. The magnetic-field maps are a data product from the Helioseismic and Magnetic Imager \citep[HMI,][]{HMI_Schou2012} onboard the Solar Dynamics Observatory \citep[SDO,][]{SDO_Pesnell2012}. Data were retrieved using the Very Fast Inversion of the Stokes Vector code \citep[VFISV,][]{VFISV_Borrero2011,VFISV_Centeno2014}. The procedure is discussed in \cite{Abbasvand2020AA}. Because the HMI Stokes $Q$ and $U$ signals were noisy in weak magnetic fields (\mbox{$|B| < 125$~G}), the inversion code returned false values of the inclination, \mbox{$\theta \approx 90^{\circ}$}. We applied a mask allowing to accept inclination values only when $|B|$ was above 125~G. The HMI data sets, including continuum intensity, magnetic-field strength, and inclination, were acquired at the middle instant of each observation. The HMI and IRIS slit-jaw continuum images were then aligned by means of a semi-automatic spatial alignment procedure described in detail in \cite{Abbasvand2020AA}. For those data sets where the continuum slit-jaw images were not recorded, we aligned the IRIS and HMI images following the heliographic coordinates of IRIS observations.


\section{Radiative cooling and deposited acoustic flux} \label{sec:Deposited}

\subsection{Model atmospheres and radiative losses}
\label{sec:Models}

We used a grid of scaled semi-empirical VAL models to assign appropriate model atmospheres to each position along the slit in each data set. The initial temperature and column-mass stratifications of the VAL C--F models \citep{Vernazza1981} were modified using the method of \cite{Abbasvand2020}. These modified stratifications serve as an input to the non-LTE radiative-transfer code based on the Multi-level Accelerated Lambda Iterations technique \citep[MALI,][]{Rybicki1991, Rybicki1992} with standard partial frequency redistribution (PRD). The code first runs its hydrogen version (with PRD in Lyman lines), where it calculates the ionization structure and populations of hydrogen levels using a five-level plus continuum atomic model, re-computes the scaled atmospheric model, and calculates the net radiative cooling rates for the hydrogen lines and continua. The microturbulent velocities are taken from the initial VAL models.
The Ca\,\textsc{ii} and Mg\,\textsc{ii} versions (with PRD in the K and H lines of Ca\,\textsc{ii} and k and h lines of Mg\,\textsc{ii}) are then run to calculate synthetic profiles of the Mg\,\textsc{ii} k and h lines and the net radiative cooling rates for Mg\,\textsc{ii} k and h, Ca\,\textsc{ii} K and H, and the Ca\,\textsc{ii} infrared triplet. A sum of all the contributions to the net radiative cooling rates, computed for each height in the model, are the total radiative losses.

Having computed the atomic level populations for all bound states and the mean radiation fields of all line transitions, we evaluate the net radiative cooling rates using a standard formula; see for example Eq. (21) of \cite{Vernazza1981}. For bound-free  hydrogen continua we integrate  the differences between emission and absorption over the continuum wavelengths; Ca\,\textsc{ii} and Mg\,\textsc{ii} continua are neglected. We tested this standard approach using the VAL C model and our loss function relatively closely resembles  that shown in Fig.~49 of \cite{Vernazza1981}. Also, our total losses integrated  over chromospheric heights are in agreement with those in Table 29 of the latter paper. We note that these losses are completely dominated by hydrogen, Ca\,\textsc{ii}, and Mg\,\textsc{ii} lines.
A grid of 2806 scaled models, synthetic profiles, and corresponding radiative losses is calculated for each initial VAL model.

The best-matching models were obtained by fitting synthetic k and h profiles to the observed time-averaged ones. We must note that the observed profiles
have broader emission cores and shallower central reversals than our pre-computed synthetic profiles. This is partly caused by the IRIS instrumental profile, which can be described by a Gaussian function with a full width at half maximum of 53~m\AA ~\citep{DePontieu2014SoPh}. However, a major part of the broadening is model-dependent. \citet{Judge2020} pointed out that the effect can be explained by 3D radiative transfer but when 1D models are used, the broadening is mimicked by micro- and/or macroturbulent motions. Because the microturbulence is fixed in our grid of models, we adopted the macroturbulence as a free parameter. The macroturbulence can be approximated by a Gaussian distribution of velocities. Therefore, the synthetic profiles were first convolved with a Gaussian kernel that included the instrumental profile and macroturbulence and were then compared with the observed ones.

\begin{figure*}[t]\centering
\includegraphics[width=0.8\textwidth]{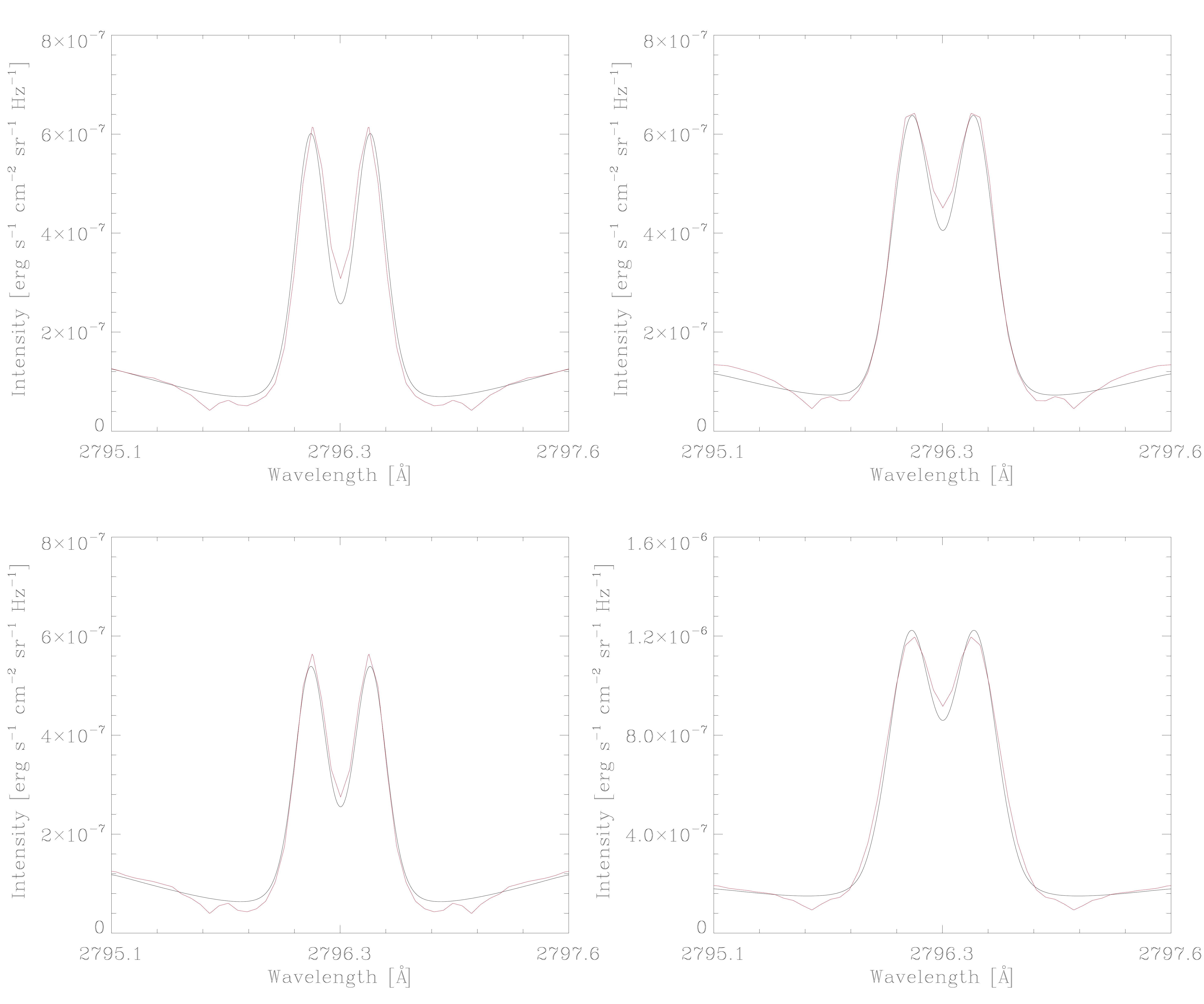}
\caption{Example fit of Mg\,\textsc{ii} k synthetic profiles ({\it black}) to the observed time-averaged ones ({\it red}) for two different positions in quiet (QS5, {\it left column}) and active (AR7, {\it right column}) regions.
\label{fig:linefit}}
\end{figure*}

The model atmospheres were assigned to different positions along the slit of each data set in two steps: (1) to find the best-matching initial VAL model and macroturbulent velocity and (2) to select the optimal final model from the grid of scaled models.
Because we do not know the height profiles of the velocity fields and the velocities themselves are small, we simply used the static code, which computed only a half of the synthetic profile, beginning at $\Delta \lambda = 0$. The entire synthetic profiles were completed by mirroring and the observed time-averaged profiles were also symmetrized.
First, synthetic profiles of four initial VAL models convolved with macroturbulent velocities characterized by Gaussian $\sigma$ in the range of 0--8~km\,s$^{-1}$ and the instrumental broadening with equivalent $\sigma \simeq$~2.41~km\,s$^{-1}$ were compared with the observed profiles to find the best match, that is, the smallest value of the sum of the squared differences between the convolved synthetic and observed time-averaged profiles (merit function).
The matching distributions of macroturbulent velocity were characterized by $\sigma = 4$--7~km\,s$^{-1}$ with a maximum at 5.5~km\,s$^{-1}$.
The matching initial VAL model and macroturbulent velocity at each position along the slit were then used as an input to the second step. Synthetic profiles produced by the grid of models derived from the initial one were convolved with the macroturbulent velocity and instrumental broadening and the final model for each position along the slit was selected by minimizing the merit function.
The fit of synthetic to time-averaged observed profiles is illustrated in Fig.~\ref{fig:linefit}, where two examples of quiet-Sun profiles ({\it left}) and two examples of active region profiles ({\it right}) are shown.

Knowledge of effective formation heights of the \mbox{Mg\,\textsc{ii} k and h} lines in the solar chromosphere is essential for determination of the range of heights at which the energy carried by acoustic waves and that released by radiation have to be compared. The formation heights can be obtained from a contribution function $C_\lambda(h)$ \citep[e.g.,][]{Gurtovenko1974} that characterizes the contribution of different atmospheric layers to the emergent intensity of radiation at the considered wavelength $\lambda$. For the whole grid of scaled models, contribution functions are computed by the non-LTE radiative-transfer code described above. The effective formation height can be calculated as a mean $h$ weighted by $C_\lambda(h)$ at a given $\Delta\lambda$ from the line center.
We have to find effective formation heights $h_1,\, h_2$ of those parts of the line profile where the Doppler velocities are measured, that is, the central reversal ($h_2, \Delta\lambda = 0$) and the half-maximum intensity of the emission core ($h_1$). Here, $\Delta\lambda$ varies in the range 0.20--0.23~\AA ~for the k line and \mbox{0.18--0.21~\AA} ~for the h line, according to the model atmosphere.
We obtained effective formation heights $h_1$ and $h_2$ for all models assigned along the slit of each data set. The Mg\,\textsc{ii} k and h lines behave similarly. The emission core of the k line is formed only a few tens of kilometers higher than the core of the h line, and so we define the effective formation heights $h_1$ and $h_2$ as the average of the heights obtained for both lines. The total radiative losses $L$ are integrated in the range between these two heights for each position along the slit in each data set.

\begin{figure}[t]\centering
\includegraphics[width=3.3in]{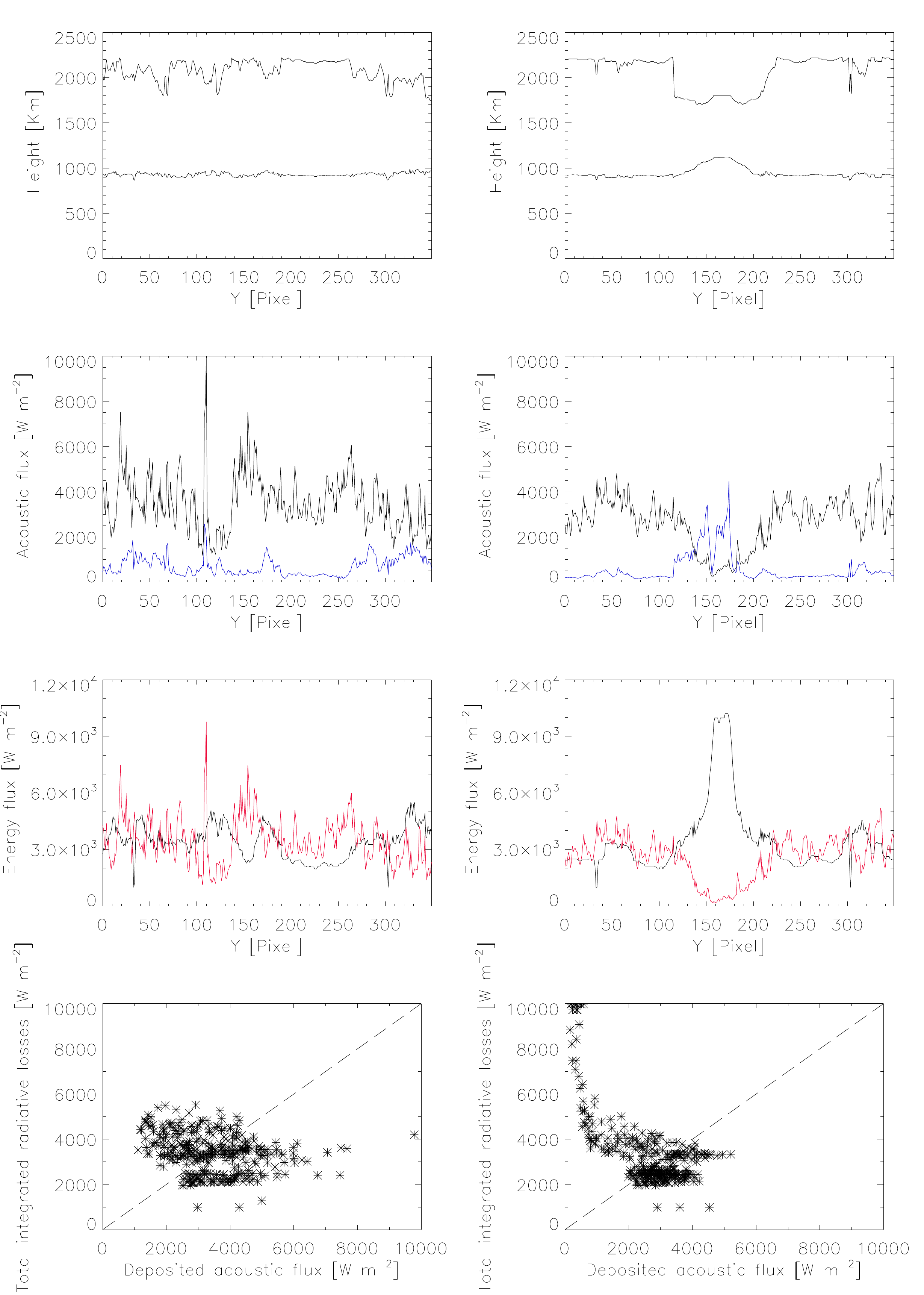}
\caption{Lower and upper effective formation heights ({\it first row}), acoustic fluxes ({\it second row}), and energy fluxes ({\it third row}) along the slit in the typical quiet-Sun data set, QS11 ({\it left}), and active-region data set, AR10 ({\it right}). In the acoustic-flux plots, black lines correspond to incoming acoustic fluxes and blue ones to outgoing acoustic fluxes multiplied by ten. In the energy-flux plots, red lines correspond to deposited acoustic fluxes and black ones to total integrated radiative losses. Scatter plots of these quantities are in the {\it fourth row}. Straight dashed lines represent the full balance of radiative losses by acoustic-flux deposit.   
\label{fig:fig3}}
\end{figure}

The top row in Fig.~\ref{fig:fig3} shows these heights in typical quiet- and active-Sun data sets (QS11 and AR10). We see that the effective formation heights of the central reversal at most of the positions in the quiet Sun, which are characterized by low-temperature model atmospheres, are between 2000 and 2200~km, while those of active regions (high-temperature models) are mainly in the range of 1700--1900~km. For the emission core, the effective formation heights are in the range 900--1000 and 1000--1200 km in quiet and active regions, respectively. The range between the formation heights of the emission-core half maximum and the central reversal corresponds to the middle and upper chromosphere.

\subsection{Acoustic flux}
\label{sec:Flux}

For each position along the slit of each data set, oscillations are studied using standard Fourier analysis of the time-series of Dopplergrams measured in the central reversals and the emission cores of the Mg\,\textsc{ii} k and h lines. The cadences of the time-series are between 21 and 37~s. This means that the maximum detectable frequencies are in the range of \mbox{14--24 mHz}. The frequency-resolution step is \mbox{0.06--0.27~mHz}, given by the lengths of the time-series. The method of \citet{Rieutord2010} is used to calculate the power spectra and calibrate them in absolute units.

Following the method of \cite{bello2009}, we estimated the acoustic-energy fluxes at two chromospheric heights $h_1$ and $h_2$ (Section~\ref{sec:Models}) at each position along the slit for each data set, assuming that the acoustic waves propagate upwards (cf. Section~\ref{sec:Wave}). The presence of inclined magnetic field observed at some locations is taken into account by setting the acoustic cut-off frequency to $\nu_{\rm ac} = 5.2 \cos \theta$~mHz, where $\theta$ is the magnetic-field inclination \citep{Cally2006}.
We assume that the inclination of magnetic field along the track of waves is equal to that measured in the photosphere.
The acoustic flux is integrated over all observable frequencies above the acoustic cut-off. A detailed description of the method is given by \cite{Abbasvand2020AA}. The acoustic energy flux $\Delta F_{\rm ac}$ deposited into the chromospheric layers between the two heights can be defined as the difference between the incoming acoustic-energy flux at the lower boundary $F_{\rm ac}(h_1)$ and the outgoing one at the upper boundary $F_{\rm ac}(h_2)$.
The uncertainty of obtained $\Delta F_{\rm ac}$ is estimated to approximately $\pm 600$~\Wmm ~(see Section~\ref{sec:Statistics}).

The properties of (magneto)acoustic waves were also investigated using the analysis of wavelet spectra. For each data set and every position along the slit we computed a wavelet spectrum for both Dopplergram series measured at the two geometrical heights $h_1, h_2$. The wavelet analysis was performed using the Python PyCWT package, which is based on \cite{Torrence1998}. In addition to the wavelet spectra using the Morlet wavelet as a mother wavelet, we also computed  the cross-wavelet spectra and the wavelet coherence comparing the Dopplergram series at the lower and upper formation heights. The cross-wavelet spectrum finds regions in time-frequency space where the time-series show high common power. The wavelet coherence on the other hand  finds regions in time-frequency space where the two time-series vary in accordance with one another, but do not necessarily have high power. The wavelet coherence  also allows us to compute the phase shifts between the series.

\section{Results} \label{sec:Results}

\subsection{Cases of typical quiet and active regions} \label{sec:typical}

Relations between the acoustic fluxes and total integrated radiative losses are shown for two data sets typical for the quiet Sun (QS11) and active region (AR10). The second row in Fig.~\ref{fig:fig3} shows the incoming acoustic fluxes $F_{\rm ac}(h_1)$ (black line) and the outgoing ones $F_{\rm ac}(h_2)$ (multiplied by ten, blue line) for each position along the slit. A fraction of the incoming energy flux is dissipated in the chromosphere between $h_{1}$ and $h_{2}$ and the rest continues to propagate higher in the atmosphere. The deposited acoustic energy fluxes $\Delta F_{\rm ac}$ (red) and the total integrated radiative losses $L$ (black) are compared in the third row. Finally, scatter plots of $L$ versus $\Delta F_{\rm ac}$ are displayed in the fourth row.

In the quiet-Sun data set, the total radiative losses integrated over the 800--1300 km thick layers are in the range $1000 < L < 5500$~\Wmm, relating to the network and intranetwork structures. The ratio of the mean deposited acoustic flux to the radiative losses $\overline{\Delta F_{\rm ac}}/L$ is between 0.9 and 1.5, meaning that for most of the positions along the slit in the plot (Fig.~\ref{fig:fig3},  third and fourth rows on the  left), the deposited acoustic flux is sufficient to balance the radiative losses and maintain the observed chromospheric temperature. In this case, 98\% of the incoming acoustic flux is absorbed between the heights $h_1$ and $h_2$ and only about 70~\Wmm ~on average passes higher, to the transition region and corona.

In the active-region data set, the plots (Fig.~\ref{fig:fig3},  third and fourth rows on the right) show that $L \gg \Delta F_{\rm ac}$ for pixels $140 < y < 200$, meaning that the energy deposited by (magneto)acoustic waves between the considered heights is very small and the ratio $\overline{\Delta F_{\rm ac}}/L$ is approximately 0.1--0.3. In this area, the integration is made over a   650~km thick layer in the central part of the plage, where $L > 4000$~\Wmm. The incoming acoustic flux is reduced here by a factor of 3.5 compared to the surrounding quiet area, and about 20\% of this flux passes the layer without being absorbed. The remaining pixels at $y < 140$ and $ y > 200$ correspond to the quiet Sun and $L$ is balanced by $\Delta F_{\rm ac}$ at these positions.

\subsection{Comparison of deposited acoustic flux with radiative losses in all data sets}
\label{sec:Statistics} 

\begin{figure}[t]\centering
\includegraphics[width=3.3in]{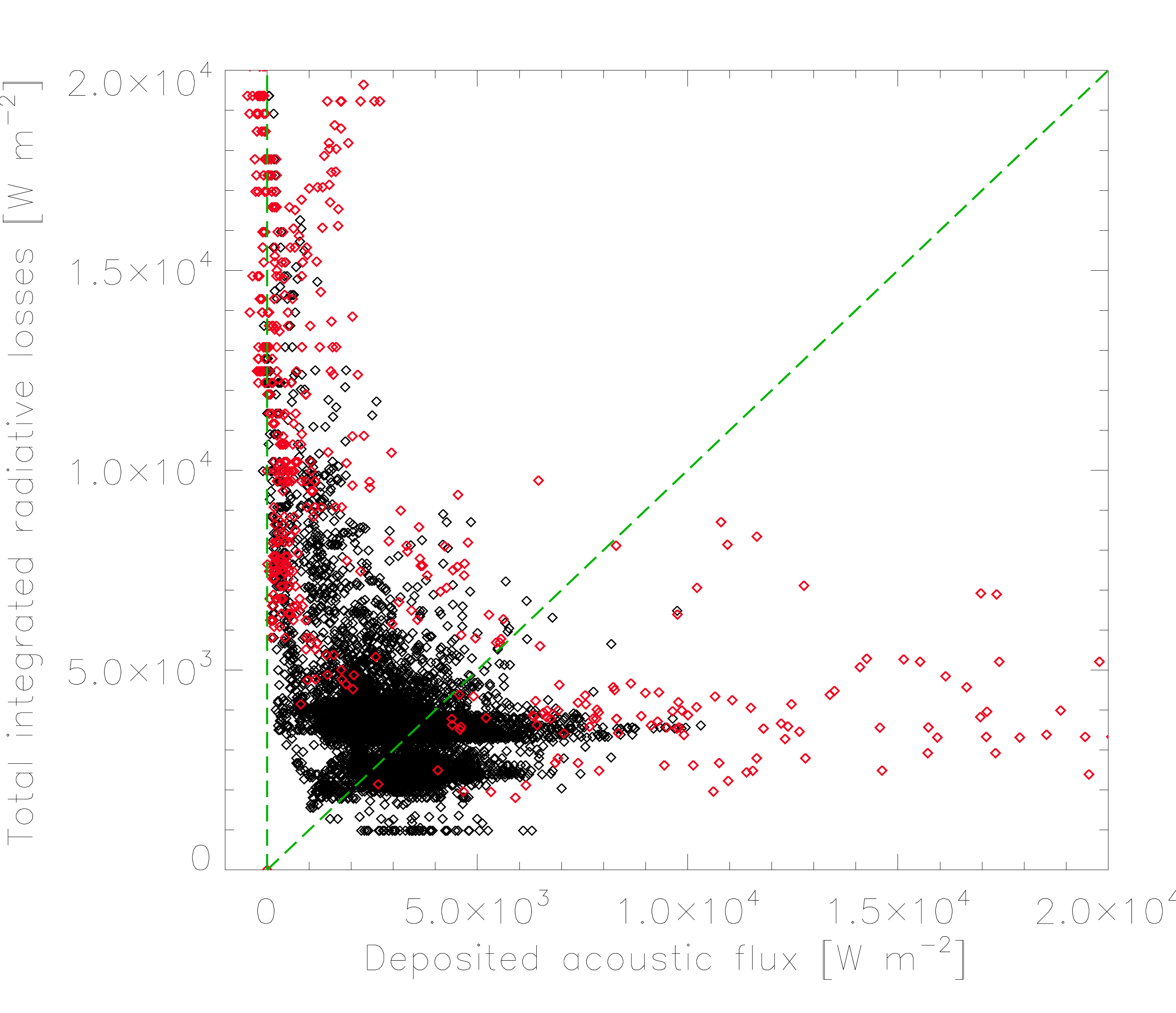}
\caption{Scatter plot of total integrated radiative losses versus deposited acoustic flux at all positions of  all data sets. Black and red diamonds correspond to nonmagnetic and magnetic positions, respectively. Straight dashed lines represent the full balance of radiative losses by acoustic-flux deposit and zero deposited acoustic flux.  
\label{fig:fig2}}
\end{figure}

In total, 23 different data sets including 12 quiet-Sun and 11 active regions were used to compare $\Delta F_{\rm ac}$ and $L$ in the middle and upper chromosphere.
The quiet-Sun atmospheres are represented by scaled initial VAL C and D models, providing the best match of synthetic to local time-averaged observed profiles. In the case of active-region atmospheres, best-matching models from the grids of scaled initial VAL E and F models were selected. The most important ingredients for $\Delta F_{\rm ac}$ calculation are the Doppler velocity power spectra, gas density and pressure at the lower- and upper-boundary heights, and the magnetic-field inclination. 
Figure~\ref{fig:fig2} illustrates the scatter-plot comparison of $\Delta F_{\rm ac}$ and $L$ for all positions along the slits in all data sets. Points located in nonmagnetic regions (6337 in total) are distinguished from those (524) of magnetic regions with $|B| > 125$~G.

Figure~\ref{fig:fig2} shows that all the observed quiet and active regions behave similarly to the examples presented in Section~\ref{sec:typical}. In quiet-Sun regions ($L < 5000$~\Wmm), $\Delta F_{\rm ac}$ is statistically comparable to $L$, meaning that the energy released by radiation can be balanced by the deposited acoustic-flux energy. Most of the incoming acoustic flux is absorbed in the studied layer and on average only 2\% of it passes to the transition region and corona.

In active regions ($L > 5000$~\Wmm), $\Delta F_{\rm ac}$ decreases with increasing $L$, supplying only 10--30\% of the radiated energy and even less, namely 5\%, for $L > 10000$~\Wmm. This decrease is caused mainly by the observed reduction of incoming acoustic flux in active regions, which is probably due to the magnetic-canopy effect. Another reason is that 10--20\% of the incoming acoustic flux is not deposited in the studied layer and escapes higher into the atmosphere (cf. Section~\ref{sec:Wave}). 

The uncertainty of obtained $\Delta F_{\rm ac}$ depends on several factors: the accuracy of the Doppler velocity measurement, correct assignment of the model atmosphere, and the accuracy of formation heights from which the density and gas pressure at the lower and upper boundaries are obtained. Therefore, we attempt to estimate this uncertainty directly from spatial fluctuation of $\Delta F_{\rm ac}$ values, assuming that they do not vary abruptly with position along the spectrograph slit at distances corresponding to the typical size of chromospheric features seen in the slit-jaw images, which is approximately 1.5$''$.  We define the lower limit of the uncertainty as the standard deviation of differences between the original values of $\Delta F_{\rm ac}$ and those smoothed by a window of 1.5$''$ in width. The upper limit is estimated as the standard deviation of differences between $\Delta F_{\rm ac}$ at neighboring positions. From six quiet-Sun data sets with low magnetic signal (QS4, QS7, and QS9--12) we find that the lower and upper limits are 520~\Wmm ~and 640~\Wmm ~respectively, meaning that the uncertainty is approximately $\pm 600$~\Wmm. The points with negative $\Delta F_{\rm ac}$ observed for $L > 12000$~\Wmm ~(Fig.~\ref{fig:fig2}) are inside this range of uncertainty.
The points with \mbox{$\Delta F_{\rm ac} > 8000$~\Wmm} seen in Fig.~\ref{fig:fig2} ~originate in most cases from magnetic regions. Their outlying values stem from errors of measurements of the k and h emission-core Doppler velocity together with an enhancement of $\Delta F_{\rm ac}$ in strongly inclined magnetic fields.

\subsection{Properties of observed waves} \label{sec:Wave}

\begin{figure}
    \centering
    \includegraphics[width=0.49\textwidth]{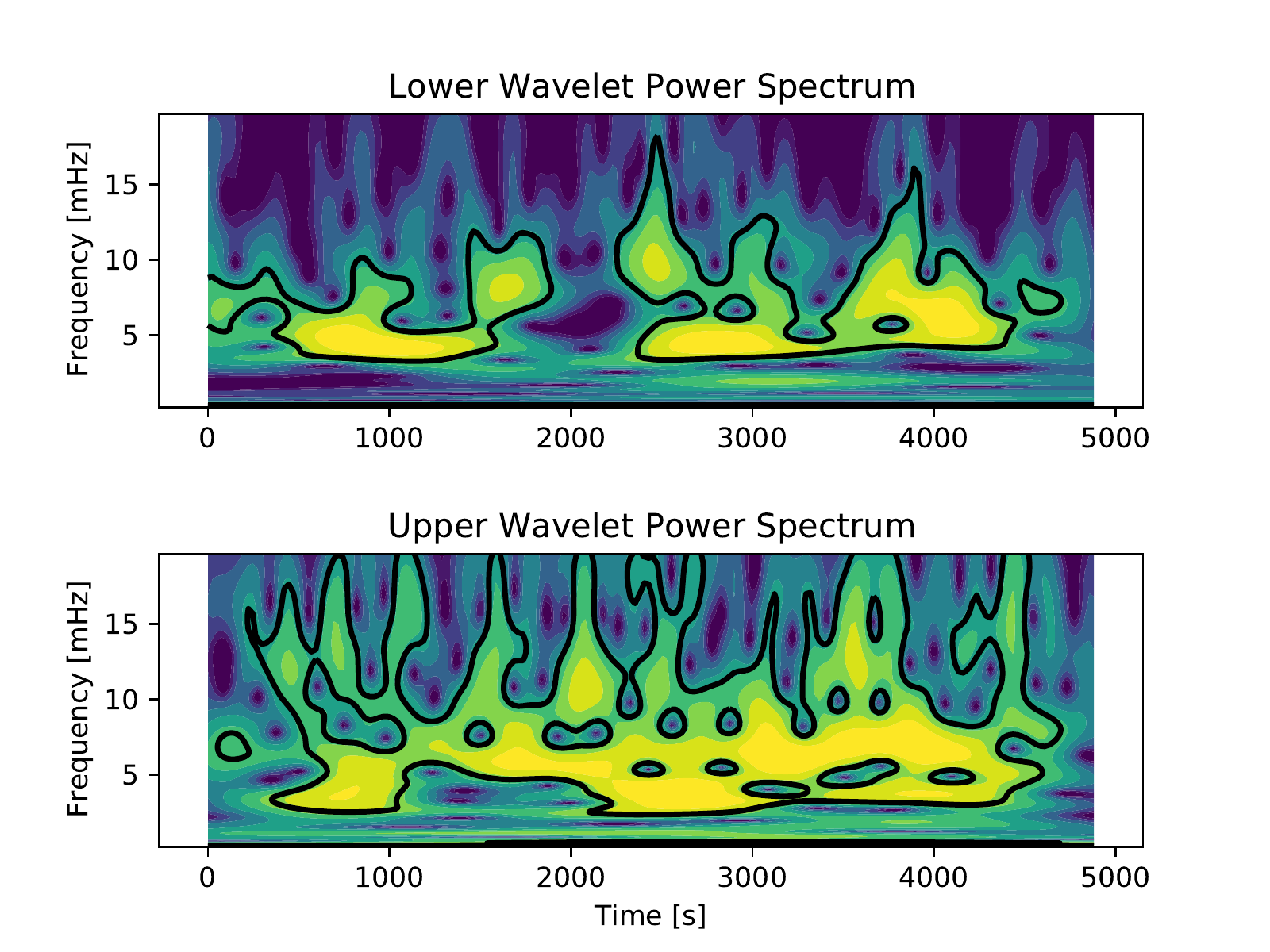}\\
    \includegraphics[width=0.49\textwidth]{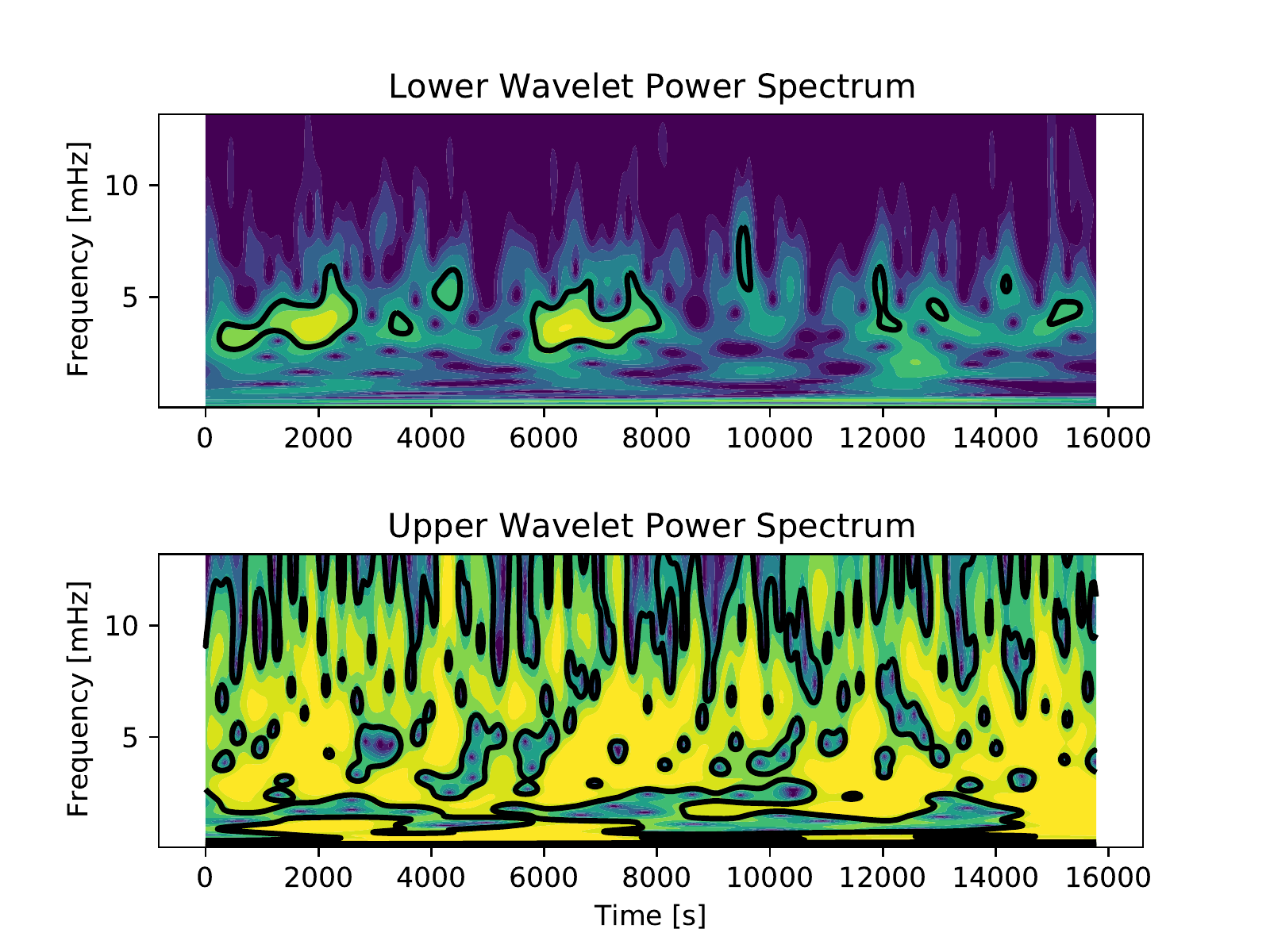}
    \caption{Examples of the wavelet power spectra computed for a pixel in the nonmagnetic atmosphere (upper pair of plots) and for a pixel classified as magnetized atmosphere (lower pair of plots), always for the lower- and upper-height boundary of the chromospheric layer under study. The solid black contours encircle the islands of statistical significance of the power.}
    \label{fig:wavelet_spectra}
\end{figure}

The wavelet spectra obtained from the Dopplergrams at each observed pixel, that is, a position along the slit in each data set, show a relatively complicated pattern (see examples in Fig.~\ref{fig:wavelet_spectra}). At the lower-height boundary, a significant power excess is seen in the frequency band around 5~mHz. The islands of the significance on the wavelet power spectrum have an intermittent nature, with a period on the order of 500--1000~s with a significant power, swapping with a period of the order of 200--500~s where the wavelet power is not above the significance threshold for the nonmagnetic (quiet Sun) pixels. For the pixels classified as a magnetized atmosphere, where the magnetic induction is above the 125~G threshold, the wavelet power seems to be shifted towards smaller frequencies and the intermittency is more severe, because, in general, the wavelet power is smaller in the magnetized atmosphere. 

In the case of the upper-height boundary and the nonmagnetic pixels, the wavelet power spectrum is much richer. We observe continuous islands of significant power from 3~mHz to 10~mHz and beyond. The intermittency increases with increasing frequency such that at frequencies greater than about 10~mHz the islands of significant and insignificant wavelet power swap every approximately 100~s. In the case of the magnetized atmosphere, the wavelet power seems to be present almost everywhere in the power map.

Cross-wavelet power and wavelet-coherence spectra show a complicated pattern as well. Instead of comparing the wavelet power spectra for each and every observed pixel, we took a statistical approach to learn about the typical or common behavior.

From the wavelet coherence determined for each pixel, we computed the resulting phase shift between the upper- and lower-height boundary of the chromospheric layer as a function of frequency. For each pixel, the phase shift was averaged in the time domain, separately for all time epochs and for those time epochs where the wavelet coherence was above the significance threshold. From the modeled atmosphere, we determined the geometrical height difference. Knowing these two quantities, we could determine the time lags between the two boundaries and the corresponding phase speeds as a function of frequency. We followed the classical wave description, taking
$\Delta\varphi=\omega \Delta t$ and $\Delta h = c\Delta t$,
where $\Delta\varphi$ stands for the determined wavelet phase shift, $\omega$ is the frequency, $\Delta t$ the time lag, $\Delta h = h_1 - h_2$ the geometrical height difference, and $c$ is the phase speed.

\begin{figure}
    \centering
    \includegraphics[width=0.49\textwidth]{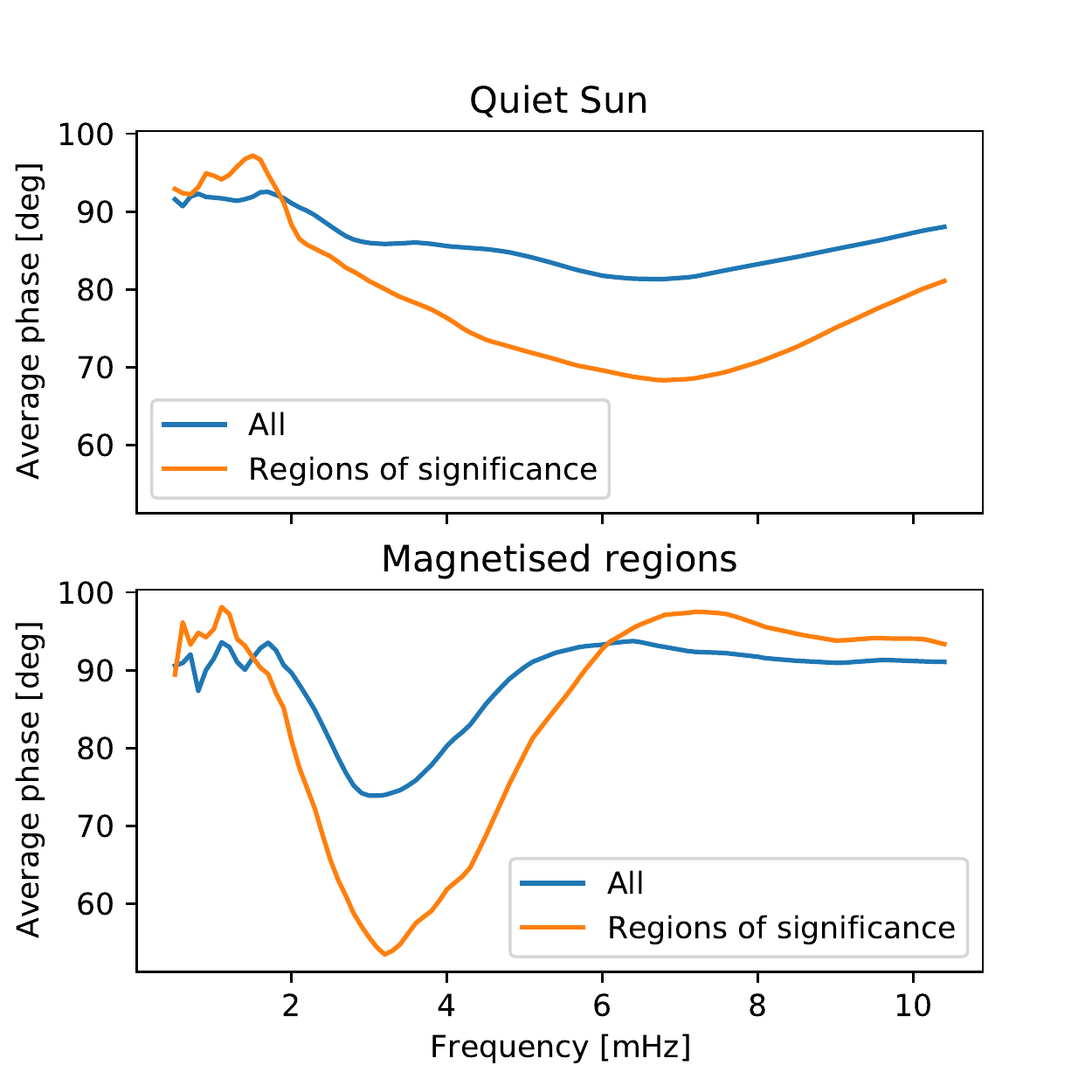}
\caption{Average wave phase shifts between the upper- and lower-height boundaries of the chromospheric layer under study plotted separately for the nonmagnetic (upper panel) and magnetic (lower panel) pixels. The averages are plotted for all contributions in the time-frequency wavelet spectrum (blue) and only for those contributions where a statistically significant wavelet coherence was indicated (orange).}
    \label{fig:mean_phase_shifts}
\end{figure}

In order to learn about the typical behavior, we averaged the phase shifts over all representations of the nonmagnetic (quiet-Sun) pixels and pixels in the magnetized atmosphere. The average phase shifts $\Delta\varphi$ computed separately for the nonmagnetic and magnetic pixels are plotted in Fig.~\ref{fig:mean_phase_shifts}. On average, we clearly see indications for the upward propagating waves (the phase shifts are positive) with typical phase shifts from 60 to 100 degrees. We have to note that we do not consider the information obtained for the frequencies smaller than 2~mHz to be trustworthy, as these frequencies are near the resolution limits 0.12--0.54~mHz given by the lengths of the observing series. The drop in the phase shift between 2~mHz and 6~mHz for the magnetic pixels appears to be real. In this frequency region, we most often observe a significant wavelet coherence in the magnetic pixels. The phase shifts correspond to the time-lags between 30 and 100~s decreasing with frequency. This characteristic time-shift was independently confirmed by a cross-correlation of the Dopplergram series in selected pixels.

\begin{figure}
    \centering
    \includegraphics[width=0.49\textwidth]{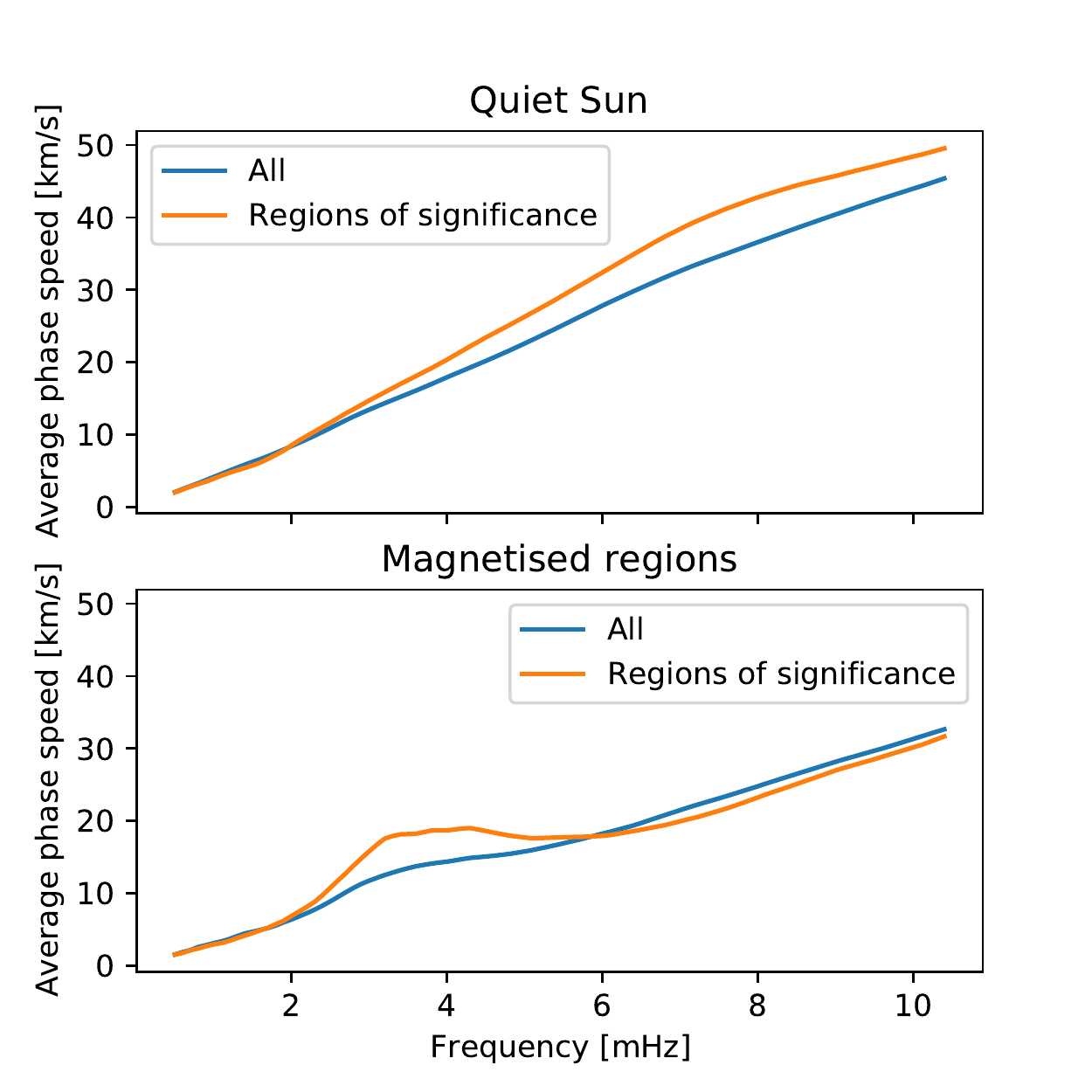}
\caption{Average wave phase speeds between the upper- and lower-height boundaries of the chromospheric layer under study plotted separately for the nonmagnetic (upper panel) and magnetic (lower panel) pixels. The averages are plotted for all contributions in the time-frequency wavelet spectrum (blue) and only for those contributions where a statistically significant wavelet coherence was indicated (orange).}
    \label{fig:mean_phase_speeds}
\end{figure}

The mean $\Delta h$ is about 1200~km for the nonmagnetic pixels and about 700~km for the magnetic ones. This allows us to infer the average phase speed corresponding to the wave propagation. These average phase speeds are plotted in Fig.~\ref{fig:mean_phase_speeds}.
According to our atmospheric models, the typical speeds of sound are \mbox{7--8 k\mps} for the lower-height boundary and 10--12~k\mps ~for the upper-height boundary. We cannot determine the Alfv\'en speed directly as we do not know the induction of the magnetic field in these layers. By taking a reasonable estimate of 100~G, we obtain estimates for the Alfv\'en speeds of 20--40~k\mps.

The Dopplergram visualization (Fig.~\ref{fig:Dop20}) indicates that the time evolution of the Doppler velocity  behaves differently at the lower  and upper height boundaries. Whereas at the lower boundary the Dopplergram series is visually compatible with the pattern of waves, at the upper boundary the wave-like character is suppressed by a structure similar to shocks; that is, the Doppler velocity exhibits a fast increase in magnitude followed by a much slower decrease. This is true for the quiet-Sun regions, but for the magnetized regions, the character of the Doppler velocity resembles waves even at the upper boundary.

The phase speeds for the regions with a large statistical significance displayed in Fig.~\ref{fig:mean_phase_speeds} indicate that the signal with frequencies higher than 2~mHz, which we find trustworthy, spreads with supersonic velocities. With the exception of the frequency range 3--5~mHz for the magnetic pixels, the phase speed monotonously increases with frequency. This is consistent with the nature of shocks, where the fast increases in Doppler velocity are naturally mapped to high frequencies in the frequency space, as the steep increases are characteristic of discontinuities.

In the frequency range 3--5~mHz for the magnetic pixels, the phase speed is almost constant, at about 18~k\mps. This is supersonic and likely corresponds to the presence of the fast magnetoacoustic mode propagating upwards. We need to mention here that the typical inclination of the magnetic field in those pixels is non-negligible but rather small on average. We therefore compare the pixels at the lower- and upper-height boundaries that do not have to share the same field line. Thus, in principle, we cannot observe pure Alfv\'enic motions. Because of the line of sight (with respect to the direction of the magnetic field) we possibly observe a component of the hydromagnetic waves, and as the derived phase speed is supersonic, we interpret this as a fast mode. At this point we cannot rule out the presence of flux-tube waves. On the other hand, the wavelet spectra do not resemble the power spectra of the wave trains indicating flux-tube waves \citep[see e.g.,][]{Meszarosova2020}.


\section{Summary and conclusions \label{sec:conclusion}}

We examined the heating of the solar chromosphere by dissipation of (magneto)acoustic  waves in the middle and upper chromosphere using IRIS spectroscopic observations in the Mg\,\textsc{ii} k and h lines. A quantitative analysis was applied to compare the deposited acoustic energy flux with the total integrated radiative losses in 12 quiet and 11 active regions located in the central zone of the solar disk. 
The deposited acoustic flux was derived from time-series of Dopplergrams measured in the central reversals and at the half-maximum intensity of the emission cores of the Mg\,\textsc{ii} k and h lines. On average, the central reversals were formed at heights of 2200~km in quiet regions and 1800~km in active regions, setting the upper boundary of the studied chromospheric layer. The lower boundary was defined by the average formation height of the emission cores, 900~km in quiet regions and 1100~km in active regions. The individual heights of these boundaries vary with the model atmosphere. The maximum detectable frequencies of velocity oscillations were in the range of 14--24~mHz. We made use of non-LTE hydrostatic semi-empirical models to calculate the radiative losses $L$, which were assigned to each position along the slit of each data set together with the deposited acoustic fluxes $\Delta F_{\rm ac}$. The models were selected by scaling the temperature and column-mass stratifications of the initial models VAL C--F to provide the best match between synthetic and observed time-averaged profiles. We compared $L$ and $\Delta F_{\rm ac}$ in the chromospheric layer delimited by the upper and lower boundaries.

We also studied properties of observed waves using a wavelet analysis of the Dopplergrams at the lower and upper boundaries. The wavelet coherence spectra provided phase shifts between the two boundaries. The average phase shifts for the nonmagnetic and magnetic regions are positive (Fig.~\ref{fig:mean_phase_shifts}), meaning that the direction of propagation of the (magneto)acoustic waves is upward through the solar chromosphere. The phase speeds of the signals with frequencies higher than 2~mHz are supersonic. The increase of phase speed with frequency in nonmagnetic regions is consistent with the presence of supersonic shocks. In magnetic regions, the phase speed is roughly constant in the frequency range 3--5~mHz, which is an indication of fast supersonic magnetoacoustic modes.

We have shown that the deposited acoustic flux can fully balance the energy released by radiation in the layers between the heights of 900 and 2200~km in quiet-Sun regions (Fig.~\ref{fig:fig2}). A dissipation of supersonic shocks can be a major contributor to the radiative losses. This result is in agreement with \cite{Abbasvand2020AA}, who reported that the acoustic flux deposited in the middle chromosphere (1000--1400~km) of a quiet region completely balanced the radiative losses, and that its contribution in the upper chromosphere above 1400~km was very small. We note that these quiet-Sun regions were far from plages and their canopies formed by extended magnetic-field lines. Otherwise, the acoustic energy flux can be reduced by a factor of 2--3 in quiet regions that are close to plages or pores \citep{Abbasvand2020} due to the presence of magnetic shadows \citep{Vecchio2007, Konto2010}.

Active regions with a strong magnetic field show a small contribution of acoustic waves, where the acoustic-flux deposit accounts for only 10--30\% of the radiative losses in the chromosphere between the heights of 1100 and 1800~km. The deposited acoustic-energy flux is small and decreases with increasing radiative losses. We hypothesize that there may be a reduction in the incoming acoustic flux due to the presence of magnetic shadows and that waves in magnetic regions, which do not dissipate as efficiently
as supersonic shocks, may behave differently. The active-region chromosphere is certainly heated by mechanisms different from the magnetoacoustic waves. This finding completes the results of \cite{Abbasvand2020AA}, who found a negligible contribution of magnetoacoustic waves to the heating of the upper chromosphere (1600--1900~km) in active regions.

The comparison of our results with the extensive literature on chromospheric heating by waves is not straightforward. Different methods with different assumptions were used to find the wave energy flux, giving different results. For example, \cite{Fossumetal2005} and \cite{Carlsson2007} applied numerical simulations to match observed broadband intensity fluctuations and concluded that the waves do not have enough energy to heat the quiet chromosphere. \cite{bello2009, Bello2010a}, as we did here, used power spectra of Doppler velocity oscillations and found that the energy deposited by the waves is sufficient.
In most previous cases, the acoustic energy flux was determined in layers below the chromosphere. Our observations allow us to calculate the acoustic energy deposited in the middle and upper chromosphere between $h \simeq 1000$ and 2000 km and compare it to radiative losses from this range of heights. We cannot include lower layers, but an important part of energy released by radiation comes from the lower chromosphere, as seen from Fig.~5 of \cite{Abbasvand2020}. The most reliable comparison should cover all heights in the chromosphere, using 3D radiative transfer and time-dependent model atmospheres.

Although our stationary approach based on time-averaged acoustic fluxes and 1D hydrostatic semi-empirical models might not be fully realistic, it nevertheless opens up the possibility of a general estimate of the role of (magneto)acoustic waves in chromospheric heating by statistical comparison of the deposited acoustic fluxes with the radiative losses. Our findings, namely (i) our discovery of the ability of the quiet-region middle chromosphere to absorb sufficient acoustic-energy flux to maintain the temperature derived from semi-empirical models and (ii) the substantial reduction of incoming and deposited acoustic fluxes in magnetized active regions, may provide important information that can be used in more realistic studies that use dynamic time-dependent models.


\begin{acknowledgements}

We thank the referee for constructive comments, which helped us to improve the paper. This work was supported by the Czech Science Foundation and Deutsche Forschungsgemeinschaft under the common grant 18-08097J---DE 787/5-1 and the institutional support ASU:67985815 of the Czech Academy of Sciences. IRIS is a NASA small explorer mission developed and operated by LMSAL with mission operations executed at NASA Ames Research center and major contributions to downlink communications funded by ESA and the Norwegian Space Centre. The SDO/HMI data are available by courtesy of NASA/SDO and the HMI science team. This research has made use of NASA's Astrophysics Data System and of the SolarSoft package for IDL.

\end{acknowledgements}


\bibliographystyle{aa}
\bibliography{bibliography1}

\begin{thebibliography}{44}
\expandafter\ifx\csname natexlab\endcsname\relax\def\natexlab#1{#1}\fi

\bibitem[{{Abbasvand} {et~al.}(2020{\natexlab{a}}){Abbasvand}, {Sobotka},
  {Heinzel}, {{\v{S}}vanda}, {Jur{\v{c}}{\'a}k}, {del Moro}, \&
  {Berrilli}}]{Abbasvand2020}
{Abbasvand}, V., {Sobotka}, M., {Heinzel}, P., {et~al.} 2020{\natexlab{a}},
  \apj, 890, 22

\bibitem[{{Abbasvand} {et~al.}(2020{\natexlab{b}}){Abbasvand}, {Sobotka},
  {{\v{S}}vanda}, {Heinzel}, {Garc{\'\i}a-Rivas}, {Denker}, {Balthasar},
  {Verma}, {Kontogiannis}, {Koza}, {Korda}, \& {Kuckein}}]{Abbasvand2020AA}
{Abbasvand}, V., {Sobotka}, M., {{\v{S}}vanda}, M., {et~al.}
  2020{\natexlab{b}}, \aap, 642, A52

\bibitem[{{Beck} {et~al.}(2009){Beck}, {Khomenko}, {Rezaei}, \&
  {Collados}}]{Beck2009}
{Beck}, C., {Khomenko}, E., {Rezaei}, R., \& {Collados}, M. 2009, \aap, 507,
  453

\bibitem[{{Bel} \& {Leroy}(1977)}]{Bel1977}
{Bel}, N. \& {Leroy}, B. 1977, \aap, 55, 239

\bibitem[{{Bello Gonz{\'a}lez} {et~al.}(2009){Bello Gonz{\'a}lez}, {Flores
  Soriano}, {Kneer}, \& {Okunev}}]{bello2009}
{Bello Gonz{\'a}lez}, N., {Flores Soriano}, M., {Kneer}, F., \& {Okunev}, O.
  2009, \aap, 508, 941

\bibitem[{{Bello Gonz{\'a}lez} {et~al.}(2010){Bello Gonz{\'a}lez}, {Flores
  Soriano}, {Kneer}, {Okunev}, \& {Shchukina}}]{Bello2010a}
{Bello Gonz{\'a}lez}, N., {Flores Soriano}, M., {Kneer}, F., {Okunev}, O., \&
  {Shchukina}, N. 2010, \aap, 522, A31

\bibitem[{{Borrero} {et~al.}(2011){Borrero}, {Tomczyk}, {Kubo},
  {Socas-Navarro}, {Schou}, {Couvidat}, \& {Bogart}}]{VFISV_Borrero2011}
{Borrero}, J.~M., {Tomczyk}, S., {Kubo}, M., {et~al.} 2011, \solphys, 273, 267

\bibitem[{{Cally}(2006)}]{Cally2006}
{Cally}, P.~S. 2006, Phil. Trans. Roy. Soc. London Ser. A, 364, 333

\bibitem[{{Carlsson} {et~al.}(2007){Carlsson}, {Hansteen}, {de Pontieu},
  {McIntosh}, {Tarbell}, {Shine}, {Tsuneta}, {Katsukawa}, {Ichimoto},
  {Suematsu}, {Shimizu}, \& {Nagata}}]{Carlsson2007}
{Carlsson}, M., {Hansteen}, V.~H., {de Pontieu}, B., {et~al.} 2007, \pasj, 59,
  S663

\bibitem[{{Carlsson} \& {Leenaarts}(2012)}]{Carlsson2012}
{Carlsson}, M. \& {Leenaarts}, J. 2012, \aap, 539, A39

\bibitem[{{Centeno} {et~al.}(2014){Centeno}, {Schou}, {Hayashi}, {Norton},
  {Hoeksema}, {Liu}, {Leka}, \& {Barnes}}]{VFISV_Centeno2014}
{Centeno}, R., {Schou}, J., {Hayashi}, K., {et~al.} 2014, \solphys, 289, 3531

\bibitem[{{Cuntz} {et~al.}(2007){Cuntz}, {Rammacher}, \&
  {Musielak}}]{Cuntz2007}
{Cuntz}, M., {Rammacher}, W., \& {Musielak}, Z.~E. 2007, \apjl, 657, L57

\bibitem[{{De Pontieu} {et~al.}(2014){De Pontieu}, {Title}, {Lemen}, {Kushner},
  {Akin}, {Allard}, {Berger}, {Boerner}, {Cheung}, {Chou}, {Drake}, {Duncan},
  {Freeland}, {Heyman}, {Hoffman}, {Hurlburt}, {Lindgren}, {Mathur}, {Rehse},
  {Sabolish}, {Seguin}, {Schrijver}, {Tarbell}, {W{\"u}lser}, {Wolfson},
  {Yanari}, {Mudge}, {Nguyen-Phuc}, {Timmons}, {van Bezooijen}, {Weingrod},
  {Brookner}, {Butcher}, {Dougherty}, {Eder}, {Knagenhjelm}, {Larsen},
  {Mansir}, {Phan}, {Boyle}, {Cheimets}, {DeLuca}, {Golub}, {Gates}, {Hertz},
  {McKillop}, {Park}, {Perry}, {Podgorski}, {Reeves}, {Saar}, {Testa}, {Tian},
  {Weber}, {Dunn}, {Eccles}, {Jaeggli}, {Kankelborg}, {Mashburn}, {Pust},
  {Springer}, {Carvalho}, {Kleint}, {Marmie}, {Mazmanian}, {Pereira}, {Sawyer},
  {Strong}, {Worden}, {Carlsson}, {Hansteen}, {Leenaarts}, {Wiesmann},
  {Aloise}, {Chu}, {Bush}, {Scherrer}, {Brekke}, {Martinez-Sykora}, {Lites},
  {McIntosh}, {Uitenbroek}, {Okamoto}, {Gummin}, {Auker}, {Jerram}, {Pool}, \&
  {Waltham}}]{DePontieu2014SoPh}
{De Pontieu}, B., {Title}, A.~M., {Lemen}, J.~R., {et~al.} 2014, \solphys, 289,
  2733

\bibitem[{{Fossum} \& {Carlsson}(2005)}]{Fossumetal2005}
{Fossum}, A. \& {Carlsson}, M. 2005, \nat, 435, 919

\bibitem[{{Freeland} \& {Handy}(1998)}]{FreelandHandy1998}
{Freeland}, S.~L. \& {Handy}, B.~N. 1998, \solphys, 182, 497

\bibitem[{{Garcia} {et~al.}(2010){Garcia}, {Klva{\v{n}}a}, \&
  {Sobotka}}]{Garcia2010}
{Garcia}, A., {Klva{\v{n}}a}, M., \& {Sobotka}, M. 2010, Central European
  Astrophysical Bulletin, 34, 47

\bibitem[{{Ghosh} {et~al.}(2019){Ghosh}, {Klimchuk}, \& {Tripathi}}]{Ghosh2019}
{Ghosh}, A., {Klimchuk}, J.~A., \& {Tripathi}, D. 2019, \apj, 886, 46

\bibitem[{{Grant} {et~al.}(2018){Grant}, {Jess}, {Zaqarashvili}, {Beck},
  {Socas-Navarro}, {Aschwanden}, {Keys}, {Christian}, {Houston}, \&
  {Hewitt}}]{Grant2018}
{Grant}, S. D.~T., {Jess}, D.~B., {Zaqarashvili}, T.~V., {et~al.} 2018, Nature
  Physics, 14, 480

\bibitem[{{Gudiksen} {et~al.}(2011){Gudiksen}, {Carlsson}, {Hansteen}, {Hayek},
  {Leenaarts}, \& {Mart{\'\i}nez-Sykora}}]{Gudiksen2011Bifrost}
{Gudiksen}, B.~V., {Carlsson}, M., {Hansteen}, V.~H., {et~al.} 2011, \aap, 531,
  A154

\bibitem[{{Gurtovenko} {et~al.}(1974){Gurtovenko}, {Ratnikova}, \& {de
  Jager}}]{Gurtovenko1974}
{Gurtovenko}, E., {Ratnikova}, V., \& {de Jager}, C. 1974, \solphys, 37, 43

\bibitem[{{Jefferies} {et~al.}(2006){Jefferies}, {McIntosh}, {Armstrong},
  {Bogdan}, {Cacciani}, \& {Fleck}}]{Jefferies2006}
{Jefferies}, S.~M., {McIntosh}, S.~W., {Armstrong}, J.~D., {et~al.} 2006,
  \apjl, 648, L151

\bibitem[{{Jess} {et~al.}(2015){Jess}, {Morton}, {Verth}, {Fedun}, {Grant}, \&
  {Giagkiozis}}]{Jess2015}
{Jess}, D.~B., {Morton}, R.~J., {Verth}, G., {et~al.} 2015, \ssr, 190, 103

\bibitem[{{Judge} {et~al.}(2020){Judge}, {Kleint}, {Leenaarts}, {Sukhorukov},
  \& {Vial}}]{Judge2020}
{Judge}, P.~G., {Kleint}, L., {Leenaarts}, J., {Sukhorukov}, A.~V., \& {Vial},
  J.-C. 2020, \apj, 901, 32

\bibitem[{{Judge} {et~al.}(2001){Judge}, {Tarbell}, \& {Wilhelm}}]{Judge2001}
{Judge}, P.~G., {Tarbell}, T.~D., \& {Wilhelm}, K. 2001, \apj, 554, 424

\bibitem[{{Kalkofen}(2007)}]{Kalkofen2007}
{Kalkofen}, W. 2007, \apj, 671, 2154

\bibitem[{{Kanoh} {et~al.}(2016){Kanoh}, {Shimizu}, \& {Imada}}]{Kanoh2016}
{Kanoh}, R., {Shimizu}, T., \& {Imada}, S. 2016, \apj, 831, 24

\bibitem[{{Kayshap} {et~al.}(2018){Kayshap}, {Murawski}, {Srivastava},
  {Musielak}, \& {Dwivedi}}]{Kayshap2018}
{Kayshap}, P., {Murawski}, K., {Srivastava}, A.~K., {Musielak}, Z.~E., \&
  {Dwivedi}, B.~N. 2018, \mnras, 479, 5512

\bibitem[{{Kontogiannis} {et~al.}(2010){Kontogiannis}, {Tsiropoula}, \&
  {Tziotziou}}]{Konto2010}
{Kontogiannis}, I., {Tsiropoula}, G., \& {Tziotziou}, K. 2010, \aap, 510, A41

\bibitem[{{Leenaarts}(2020)}]{Leenaarts2020}
{Leenaarts}, J. 2020, Living Reviews in Solar Physics, 17, 3

\bibitem[{{Leenaarts} {et~al.}(2013){Leenaarts}, {Pereira}, {Carlsson},
  {Uitenbroek}, \& {De Pontieu}}]{Leenaarts2013ApJ_formation}
{Leenaarts}, J., {Pereira}, T.~M.~D., {Carlsson}, M., {Uitenbroek}, H., \& {De
  Pontieu}, B. 2013, \apj, 772, 90

\bibitem[{{M{\'e}sz{\'a}rosov{\'a}} \&
  {G{\"o}m{\"o}ry}(2020)}]{Meszarosova2020}
{M{\'e}sz{\'a}rosov{\'a}}, H. \& {G{\"o}m{\"o}ry}, P. 2020, \aap, 643, A140

\bibitem[{{Pesnell} {et~al.}(2012){Pesnell}, {Thompson}, \&
  {Chamberlin}}]{SDO_Pesnell2012}
{Pesnell}, W.~D., {Thompson}, B.~J., \& {Chamberlin}, P.~C. 2012, \solphys,
  275, 3

\bibitem[{{Rieutord} {et~al.}(2010){Rieutord}, {Roudier}, {Rincon}, {Malherbe},
  {Meunier}, {Berger}, \& {Frank}}]{Rieutord2010}
{Rieutord}, M., {Roudier}, T., {Rincon}, F., {et~al.} 2010, \aap, 512, A4

\bibitem[{{Rybicki} \& {Hummer}(1991)}]{Rybicki1991}
{Rybicki}, G.~B. \& {Hummer}, D.~G. 1991, \aap, 245, 171

\bibitem[{{Rybicki} \& {Hummer}(1992)}]{Rybicki1992}
{Rybicki}, G.~B. \& {Hummer}, D.~G. 1992, \aap, 262, 209

\bibitem[{{Schou} {et~al.}(2012){Schou}, {Scherrer}, {Bush}, {Wachter},
  {Couvidat}, {Rabello-Soares}, {Bogart}, {Hoeksema}, {Liu}, {Duvall}, {Akin},
  {Allard}, {Miles}, {Rairden}, {Shine}, {Tarbell}, {Title}, {Wolfson},
  {Elmore}, {Norton}, \& {Tomczyk}}]{HMI_Schou2012}
{Schou}, J., {Scherrer}, P.~H., {Bush}, R.~I., {et~al.} 2012, \solphys, 275,
  229

\bibitem[{{Sobotka} {et~al.}(2016){Sobotka}, {Heinzel}, {{\v{S}}vanda},
  {Jur{\v{c}}{\'a}k}, {del Moro}, \& {Berrilli}}]{sobotka2016}
{Sobotka}, M., {Heinzel}, P., {{\v{S}}vanda}, M., {et~al.} 2016, \apj, 826, 49

\bibitem[{{Testa} {et~al.}(2014){Testa}, {De Pontieu}, {Allred}, {Carlsson},
  {Reale}, {Daw}, {Hansteen}, {Martinez-Sykora}, {Liu}, {DeLuca}, {Golub},
  {McKillop}, {Reeves}, {Saar}, {Tian}, {Lemen}, {Title}, {Boerner},
  {Hurlburt}, {Tarbell}, {Wuelser}, {Kleint}, {Kankelborg}, \&
  {Jaeggli}}]{Testa2014}
{Testa}, P., {De Pontieu}, B., {Allred}, J., {et~al.} 2014, Science, 346,
  1255724

\bibitem[{{Torrence} \& {Compo}(1998)}]{Torrence1998}
{Torrence}, C. \& {Compo}, G.~P. 1998, Bulletin of the American Meteorological
  Society, 79, 61

\bibitem[{{Ulmschneider} \& {Musielak}(2003)}]{Ulmsch2003}
{Ulmschneider}, P. \& {Musielak}, Z. 2003, in ASP Conf. Ser., Vol. 286, Current
  Theoretical Models and Future High Resolution Solar Observations: Preparing
  for ATST, ed. A.~A. {Pevtsov} \& H.~{Uitenbroek}, 363

\bibitem[{{Vecchio} {et~al.}(2007){Vecchio}, {Cauzzi}, {Reardon}, {Janssen}, \&
  {Rimmele}}]{Vecchio2007}
{Vecchio}, A., {Cauzzi}, G., {Reardon}, K.~P., {Janssen}, K., \& {Rimmele}, T.
  2007, \aap, 461, L1

\bibitem[{{Vernazza} {et~al.}(1981){Vernazza}, {Avrett}, \&
  {Loeser}}]{Vernazza1981}
{Vernazza}, J.~E., {Avrett}, E.~H., \& {Loeser}, R. 1981, \apjs, 45, 635

\bibitem[{{Wedemeyer-B{\"o}hm} {et~al.}(2007){Wedemeyer-B{\"o}hm}, {Steiner},
  {Bruls}, \& {Rammacher}}]{Wedemeyer2007}
{Wedemeyer-B{\"o}hm}, S., {Steiner}, O., {Bruls}, J., \& {Rammacher}, W. 2007,
  in Astronomical Society of the Pacific Conference Series, Vol. 368, The
  Physics of Chromospheric Plasmas, ed. P.~{Heinzel}, I.~{Dorotovi{\v{c}}}, \&
  R.~J. {Rutten}, 93

\bibitem[{{W{\"u}lser} {et~al.}(2018){W{\"u}lser}, {Jaeggli}, {De Pontieu},
  {Tarbell}, {Boerner}, {Freeland}, {Liu}, {Timmons}, {Brannon}, {Kankelborg},
  {Madsen}, {McKillop}, {Prchlik}, {Saar}, {Schanche}, {Testa}, {Bryans}, \&
  {Wiesmann}}]{Wulser2018SoPh}
{W{\"u}lser}, J.~P., {Jaeggli}, S., {De Pontieu}, B., {et~al.} 2018, \solphys,
  293, 149

\end{thebibliography}

\end{document}